\documentclass[a4paper, prl, twocolumn,superscriptaddress,showpacs]{revtex4}

\usepackage{amssymb}
\usepackage{amsmath}
\usepackage{epsfig}
\usepackage{color}
\usepackage{graphics, graphicx}
\usepackage{bbold}
\usepackage{psfrag}
\usepackage{mathcomp}
\usepackage{subfigure}
\usepackage{verbatim}
\usepackage{color}
\usepackage[colorlinks,citecolor=blue]{hyperref}
\def\cp#1{\mathbf{#1}}

\begin{document}

\date{\today}
\title{Three-component Ultracold Fermi Gases with Spin-Orbit Coupling}
\author{Lihong Zhou}
\affiliation{Beijing National Laboratory for Condensed Matter Physics, Institute of Physics, Chinese Academy of Sciences, Beijing 100190, China}
\author{Xiaoling Cui}
\email{xlcui@iphy.ac.cn}
\affiliation{Beijing National Laboratory for Condensed Matter Physics, Institute of Physics, Chinese Academy of Sciences, Beijing 100190, China}
\author{Wei Yi}
\email{wyiz@ustc.edu.cn}
\affiliation{Key Laboratory of Quantum Information, University of Science and Technology of China, CAS, Hefei, Anhui, 230026, People's Republic of China}
\affiliation{Synergetic Innovation Center of Quantum Information and Quantum Physics, University of Science and Technology of China, Hefei, Anhui 230026, China}

\begin{abstract}
We investigate the pairing physics in a three-component Fermi-Fermi mixture, where a few fermionic impurities are immersed in a non-interacting two-component Fermi gas with synthetic spin-orbit coupling (SOC), and interact attractively with one spin species in the Fermi gas. Due to the interplay of SOC and spin-selective interaction, the molecular state intrinsically acquires a non-zero center-of-mass momentum, which results in a new type of Fulde-Ferrell (FF) pairing in spin-orbit coupled Fermi systems. The existence of the Fermi sea can also lead to the competition between FF-like molecular states with different center-of-mass momenta, which corresponds to a first-order transition between FF phases in the thermodynamic limit. As the interaction strength is tuned, a polaron-molecule transition occurs in the highly imbalanced system, where the boundary varies non-monotonically with SOC parameters and gives rise to the reentrance of polaron states. The rich physics in this system can be probed using existing experimental techniques.
\end{abstract}

\maketitle

\emph{Introduction}.--
The realization of synthetic spin-orbit coupling (SOC) in ultracold atomic gases has triggered a great amount of experimental interests~\cite{gauge2exp,shuaiexp1,shuaiexp2,fermisocexp1,fermisocexp2,fermisocexp3,zhairev}. In ultracold Fermi gases, various forms of SOC in different dimensions can give rise to a wealth of exotic superfluid phases~\cite{sarma07,sato,shenoyrashba,gongzhang,soc4,soc6,iskin,thermo,2d2,2d1,melo,wmliu,helianyi,wy2d,xiaosen,puhan3dsoc,xiangfa3dsoc,xiongjunrashba}. In particular, recent studies have shown that by deforming the Fermi surface in the presence of SOC, pairing states with non-zero center-of-mass (CoM) momentum, the so-called Fulde-Ferrell (FF) states, can be stabilized over a large parameter region~\cite{fflo,michaeli-12,shenoy,puhantwobody,chuanweifflo,wyfflo,cchen,topofflo1,topofflo2,topofflo3}.
In all these cases, SOC proves to be a powerful tool of quantum control, which, when combined with the outstanding tunability of ultracold atomic gases, can often lead to novel pairing superfluidity.

\begin{figure}
\includegraphics[width=7.5cm]{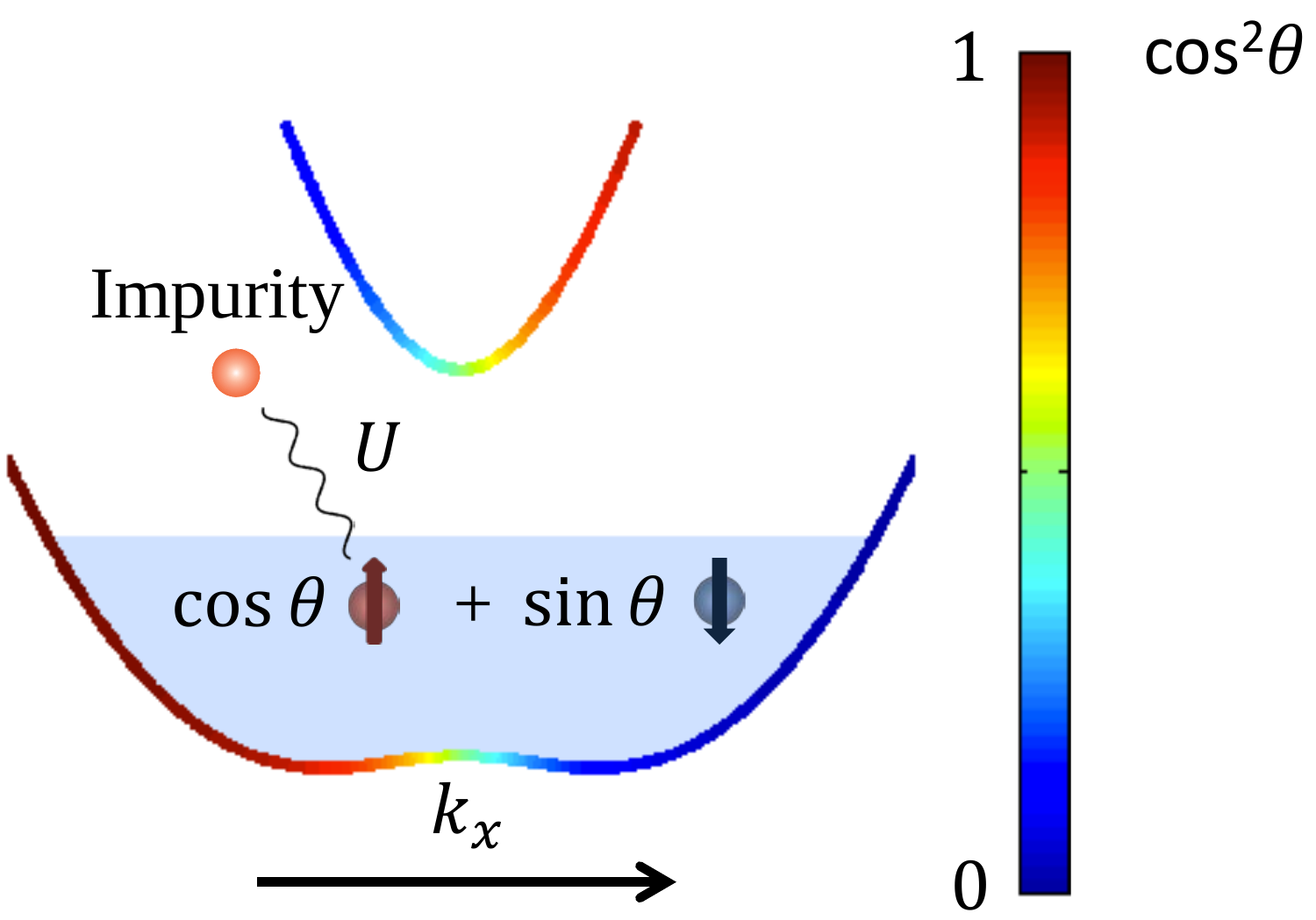}
\caption{(Color online) Schematics of the three-component Fermi-Fermi mixture. The impurity atoms interact spin-selectively with a spin-orbit coupled two-component Fermi gas. The spin superpositions in both helicity branches are momentum-dependent, as characterized by $\theta_{k}$ (see text). The pairing states naturally acquire a non-zero CoM momentum in the system.}
\label{fig:schematic}
\end{figure}

In this work, we consider another promising system for exotic pairing states, where a few fermionic impurities are immersed in a non-interacting spin-$\frac{1}{2}$ Fermi gas with the synthetic SOC that has been realized in cold atoms experiments~\cite{gauge2exp,shuaiexp1,shuaiexp2,fermisocexp1,fermisocexp2,fermisocexp3}.
In this three-component mixture, the impurity fermions are tuned close to a wide Feshbach resonance with one particular spin species in the two-component Fermi gas (see Fig.~\ref{fig:schematic}). We demonstrate that
such a system exhibits interesting pairing physics that can be readily probed using the existing experimental techniques.

A fundamental feature of the system is that the two-body bound state naturally acquires a non-zero CoM momentum, corresponding to an FF pairing phase in the thermodynamic limit. This can be attributed to the interplay between SOC and spin-selective interaction, a mechanism different from that of the previously studied FF phases in spin-orbit coupled systems~\cite{michaeli-12,shenoy,puhantwobody,chuanweifflo,wyfflo,cchen,topofflo1,topofflo2,topofflo3}.
Moreover, the existence of the spin-orbit coupled Fermi sea can lead to the competition between two FF-like molecules with different CoM momenta. Consequently, we find a first-order transition between two FF pairing phases on the mean-field phase diagram in the thermodynamic limit.
For a highly imbalanced system with a single impurity, the ground state can undergo polaron-molecule transitions, corresponding to phase transitions from the normal to the superfluid phase in many-body systems~\cite{chevy,combescot,pethick,parish,mzpolaron,koehlexp,wypolaron}.
Interestingly, it is found that the critical interaction strength for such transitions varies non-monotonically with SOC parameters, giving rise to the reentrance of polaron states for certain interaction strengths. These properties can be directly probed in Fermi-Fermi mixtures of $^{40}$K-$^{40}$K-$^{40}$K or $^{6}$Li-$^{40}$K-$^{40}$K using existing experimental techniques \cite{mzpolaron,koehlexp}.

\emph{Model}.--
The Hamiltonian of our system is written as:
\begin{align}
H=&\sum_{\cp k,\sigma}\epsilon^a_{\cp k}a^{\dag}_{\cp k,\sigma}a_{\cp k,\sigma}+\sum_{\cp k}h\left(a^{\dag}_{\cp k,\uparrow}a_{\cp k,\downarrow} +a^{\dag}_{\cp k,\downarrow}a_{\cp k,\uparrow}\right) \nonumber\\
&+\sum_{\cp k}\epsilon^b_{\cp k}b^{\dag}_{\cp k}b_{\cp k} +\frac{U}{V}\sum_{\cp k,\cp k',\cp q}a^{\dag}_{\frac{\cp q}{2}+\cp k,\uparrow}b^{\dag}_{\frac{\cp q}{2}-\cp k}b_{\frac{\cp q}{2}-\cp k'}a_{\frac{\cp q}{2}+\cp k',\uparrow}\nonumber\\
&+\sum_{\cp k}\left(\alpha k_x a^{\dag}_{\cp k,\uparrow}a_{\cp k,\uparrow}-\alpha k_x a^{\dag}_{\cp k,\downarrow}a_{\cp k,\downarrow}\right),
\end{align}
where $a_{\cp k,\sigma}$ ($\sigma=\uparrow,\downarrow$) is the annihilation operator for the spin components in the Fermi gas, and $b_{\cp k}$ is the annihilation operator for the impurity atoms. The kinetic energy of the atoms are given as $\epsilon^a_{\cp k}=\hbar^2 k^2/(2m_a)$, $\epsilon^b_{\cp k}=\hbar^2 k^2/(2m_b)$. The SOC parameters $h$ and $\alpha$ are respectively proportional to the Raman coupling strength and the momentum transfer in the Raman process generating the SOC~\cite{gauge2exp}. For simplicity, we have assumed a vanishing two-photon detuning. The bare interaction rate $U$ between the impurity and the spin-up atom can be renormalized as $1/U=1/U_p-(1/V)\sum_{\mathbf{k}}\frac{1}{(1+\eta)\epsilon^a_{\mathbf{k}}}$, where $V$ is the quantization volume, $\eta=m_a/m_b$ is the mass ratio, $U_p=2\pi\hbar^2a_s(1+\eta)/m_a$ and $a_s$ is the $s$-wave scattering length.

For a natural description of the system, we transform the annihilation operators into the helicity basis:
$a_{\cp k,\pm}=\cos\theta_{\bf k}a_{\cp k,\uparrow}+\sin\theta_{\bf k}a_{\cp k,\downarrow}$, with $\cos\theta_{\bf k}=\pm\beta_{\cp k}^{\pm}, \ \sin\theta_{\bf k}=\beta_{\cp k}^{\mp}$, and $\beta^{\pm}_{\cp k}=\big[\sqrt{h^2+\alpha^2k_x^2}\pm \alpha k_x \big]^{1/2}/\sqrt{2} [ h^2+\alpha^2k_x^2]^{1/4}$. The single-particle dispersion of the helicity branches $\xi^a_{\cp k,\pm}=\epsilon^a_{\cp k}\pm\sqrt{h^2+\alpha^2k_x^2}$ (see Fig.~\ref{fig:schematic}). The momentum dependence of $\theta_{\bf k}$ also gives rise to a momentum-dependent effective interaction between impurity atoms and atoms in the helicity branches~\cite{supp}.

\emph{Two-body states}.--
We first consider two-body bound states, which already provides key information of pairing states in the system. We adopt the ansatz wave function:
\begin{align}
\left|\psi_{\cp Q}\right\rangle=\sum_{\lambda=\pm}\sum_{\cp k}\varphi_{\cp k,\lambda}b^{\dag}_{\cp Q-\cp k}a^{\dag}_{\cp k,\lambda}\left|0\right\rangle,
\end{align}
where $\varphi_{\cp k,\pm}$ are variational coefficients, ${\cp Q}$ is the CoM momentum of the dimer, and $|0\rangle$ is the vacuum sate. The self-consistent equation for the dimer energy $E_2$ can be obtained by minimizing the energy functional $\left\langle\psi_{\cp Q}| H-E_2|\psi_{\cp Q}\right\rangle$:
\begin{equation}
\frac{1}{U}=\sum_{\lambda=\pm}\sum_{\cp k}\frac{(\beta^{\lambda}_{\cp k})^2}{E_2-\epsilon^b_{\cp Q-\cp k}-\xi^a_{\cp k,\lambda}}.\label{eqn:twobodyeqn}
\end{equation}
The ground state lies with the ${\cp Q}$-sector that gives the lowest energy solution of Eq. (\ref{eqn:twobodyeqn}). We find numerically that this always occurs for ${\cp Q}$ along the $x$-direction. For the two-body calculations, we take the unit of energy $E_0=2m_a\alpha^2/\hbar^2$, and the unit of momentum $k_0=2m_a\alpha/\hbar^2$.

\begin{figure}[tbp]
\includegraphics[width=9cm]{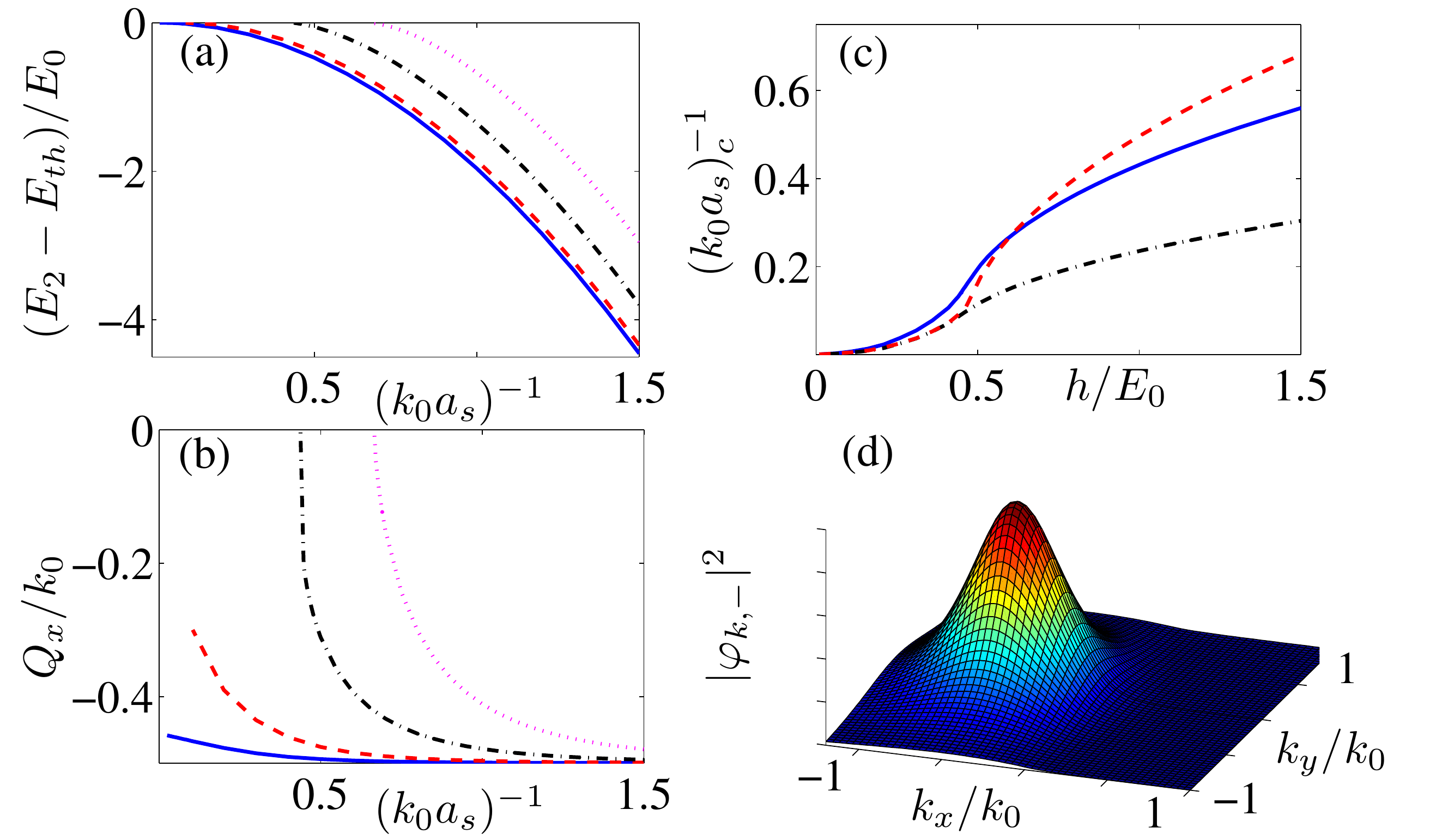}
\caption{(Color online) (a) Two-body bound state energies as functions of interaction strength for $\eta=1$, with $h/E_0=0.2$ (solid), $h/E_0=0.4$ (dashed), $h/E_0=1$ (dash-dotted), and $h/E_0=2$ (dotted). (b) Center-of-mass momentum of bound states in (a). (c) Two-body bound state threshold for varying $h$ and a fixed $\alpha k_0/E_0=1$, with the mass ratio $\eta=1$ (solid), $\eta=6/40$ (dashed), and $\eta=40/6$ (dash-dotted). (d) Typical probability distribution of the two-body wave function in the lower helicity branch, $|\varphi_{\cp k,-}|^2$, in the $k_z=0$ plane, with $\alpha k_0/E_0=1$, $h/E_0=0.2$, $(k_0 a_s)^{-1}=0.5$, and $\eta=1$. The threshold energy $E_{th}$ and the unit of energy $E_0$ are defined in the text.}
\label{fig:twobody}
\end{figure}

In Fig.~\ref{fig:twobody}(a)(b), we show the two-body ground state energy $E_2$ and the CoM momentum $Q_x$ as functions of the interaction strength with $\eta=1$. The bound state emerges when $E_2$ is below the threshold energy $E_{th}$ for two free particles~\cite{footnote_threshold}. The critical interaction strength for its emergence, $(k_0a_s)^{-1}_c$, is plotted as functions of $h$ in Fig.~\ref{fig:twobody}(c). Apparently, the presence of synthetic SOC pushes the critical $1/a_s$ toward the BEC limit, thus suppressing the formation of two-body bound state. We have also checked that three-body bound states are always metastable in this case, in contrast to the case of a Rashba SOC~\cite{cuiyi}.

A general feature of the bound state in our system is that it acquires a finite CoM momentum (Fig.~\ref{fig:twobody}(b)).
Accordingly, the momentum distribution of the bound state exhibits a peak at finite momentum as shown in Fig. ~\ref{fig:twobody}(d). Different from the finite-momentum pairing in previous studies under SOC and single-particle dispersion asymmetry~\cite{michaeli-12,shenoy,puhantwobody,chuanweifflo,wyfflo,cchen,topofflo1,topofflo2,topofflo3}, here there is no asymmetry in the single-particle dispersion. Instead, the finite-momentum pairing in our system is a combined effect of the spin-selective interaction and the momentum-dependent spin mixture under SOC, as schematically shown in Fig.~\ref{fig:schematic}. In the BEC limit, the CoM momentum approaches $-k_0/2$ (see Fig.~\ref{fig:twobody}(b)), which is effectively the momentum shift of the spin-up atom, as the system is then dominated by the binding of the impurity atom and the spin-up atom. An important implication of these two-body results is that a new type of FF-like pairing state should exist on the many-body level.

\emph{Effects of Fermi-sea}.--
En route to characterizing the many-body pairing in our system, we now consider a single impurity atom interacting with a Fermi-sea of helicity atoms, and study the effects of Fermi-sea on the pairing physics. In this highly imbalanced system, a polaron-molecule transition is expected to occur, which could be observed experimentally using radio-frequency (rf) spectroscopy~\cite{mzpolaron,koehlexp,dsjinarp}. In the thermodynamic limit, the transition should correspond to a superfluid to normal state phase transition.

\begin{figure}[tbp]
\includegraphics[width=9cm]{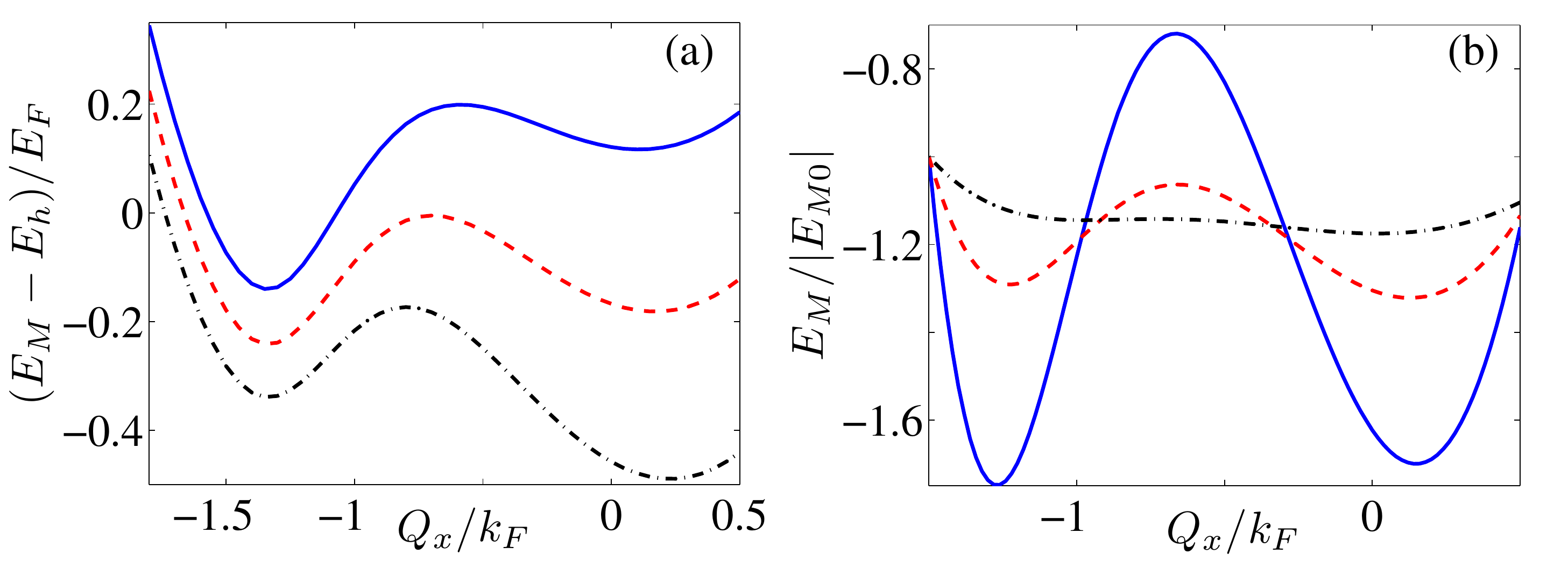}
\caption{(Color online) (a) Molecular energies relative to the Fermi surface $E_h$ as functions of the center-of-mass momentum $Q_x$ for $(k_Fa_s)^{-1}=0.5$, with $h/E_F=1.2$ (solid), $h/E_F=1$ (dashed), and $h/E_F=0.8$ (dash-dotted). (b) Normalized molecular energies as functions of $Q_x$ for $h/E_F=1$, with $(k_Fa_s)^{-1}=0.57$ (solid), $(k_Fa_s)^{-1}=0.6$ (dashed), and $(k_Fa_s)^{-1}=0.7$ (dash-dotted). The normalization in (b) is with respect to the molecular energies $E_{M0}$ at $Q_x/k_F=-1.5$. For both figures, $\alpha k_F/E_F=0.5$, and the mass ratio $\eta=40/6$ corresponds to the case of a Li-K-K mixture.
}
\label{fig:molEmvsQ}
\end{figure}

\emph{(1) Molecular state.}--
The ansatz wave function of the molecular state can be written as:

\begin{align}
\left|M_{\cp Q}\right\rangle=\sum_{\lambda=\pm}\sum_{\xi^a_{{\bf k},\lambda}>E_h}\phi_{\cp k,\lambda}b^{\dag}_{\cp Q-\cp k}a^{\dag}_{\cp k,\lambda}\left|FS\right\rangle_{N-1}, \label{mol_wf}
\end{align}
where $\phi_{\cp k,\pm}$ is the variational coefficient, $|FS\rangle_{N-1}$ denotes a spin-orbit coupled Fermi sea with $N-1$ atoms and a Fermi energy of $E_h$~\cite{supp}.
 The summation runs over the energy space beyond the Fermi surface. Minimizing the energy functional,
we can get the self-consistent equation for the molecule energy $E_M$, which is the same as Eq. (\ref{eqn:twobodyeqn}) except for the restrained summation as in Eq. (\ref{mol_wf}). We find numerically that the ground state always has a CoM ${\cp Q}$ along the $x$-direction. For the numerical calculations, we adopt the unit of energy as the Fermi energy $E_F$ of a two-component, non-interacting Fermi gas in the absence of SOC and with the same total number density. For simplicity, we only consider cases with $(\alpha k_F/E_F)^2\leq 2h/E_F$, for which there is no single-particle ground state degeneracy in the lower helicity branch. Here the Fermi wave vector $k_F$ is defined through $E_F=\hbar^2k_F^2/(2m_a)$.

The presence of the Fermi sea has significant impacts on the molecular state. Due to the Pauli blocking of atom-scattering inside the Fermi sea, we find that the molecular state becomes more difficult to form compared to the pure two-body case, and typically requires a stronger critical interaction strength. More importantly, the presence of Fermi surfaces offer possibilities of pairing at different CoM momenta, with dominating contributions from either the lower or the upper helicity branch, which can respectively give rise to negative or positive pairing momentum along $k_x$ (see Fig.~\ref{fig:schematic}).

Indeed, we find that for appropriate parameters, especially for a light impurity atom with a large $\eta$, there exists a competition between two FF-like molecular states with different CoM momenta. As shown in Fig.~\ref{fig:molEmvsQ}, first-order transitions between different FF-like molecular states can occur with varying $h$ or $(k_Fa_s)^{-1}$, while the number of atoms in the Fermi sea is fixed by $n=(1/V)\sum_{\lambda, \xi^a_{{\bf k},\lambda}<E_h}$~\cite{supp}. This strongly suggests a first-order transition between different FF pairing phases in the thermodynamic limit.

\begin{figure}[tbp]
\includegraphics[width=9cm]{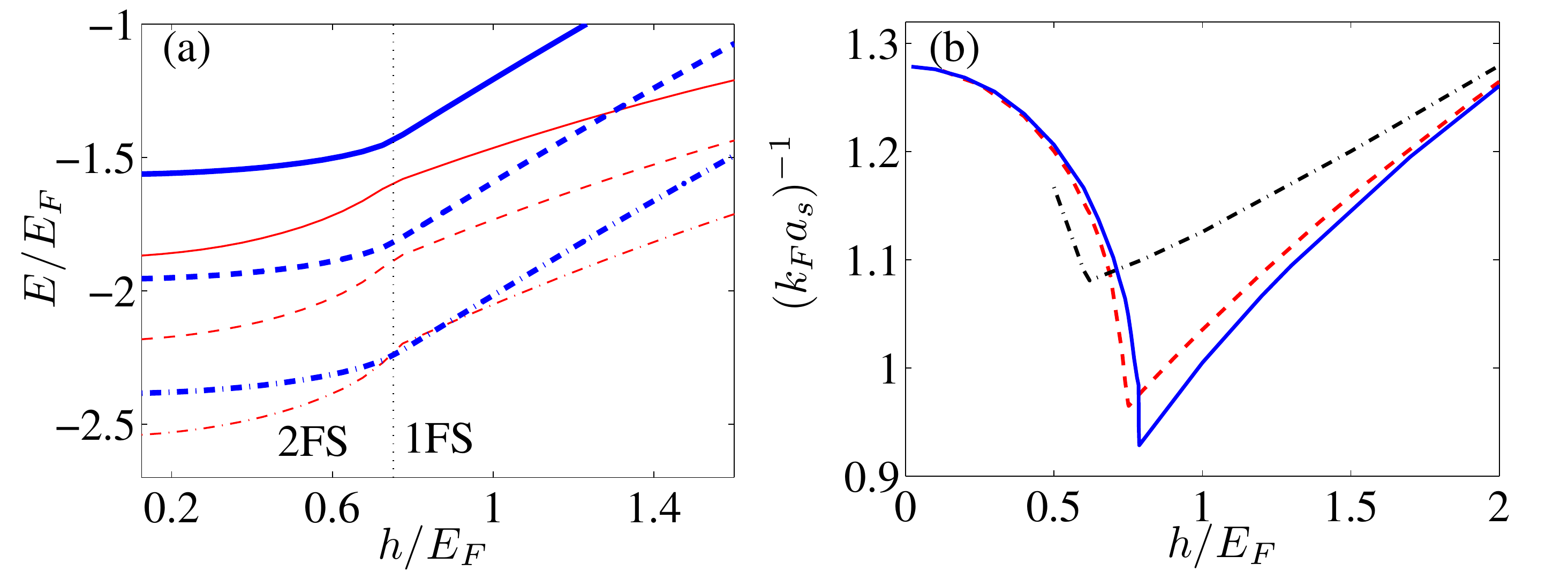}
\caption{(Color online) (a) Comparison of molecular and polaron energies for a fixed $\alpha k_F/E_F=0.5$, and different $(k_Fa_s)^{-1}$, where $(k_Fa_s)^{-1}=0.8$ (solid), $(k_Fa_s)^{-1}=0.9$ (dashed), and $(k_Fa_s)^{-1}=1$ (dash-dotted). {
The thin red curves represent polaron energy $E_p/E_F$, while the thick blue curves represent the molecular energy relative to the Fermi surface $(E_M-E_h)/E_F$.} The vertical dotted line marks the boundary between parameter regions with one Fermi surface (1FS) and with two Fermi surfaces (2FS). (b) Boundary of polaron-molecule transition, with $\alpha k_F/E_F=0.2$ (solid), $\alpha k_F/E_F=0.5$ (dashed), $\alpha k_F/E_F=1$ (dash-dotted). In all cases, the mass ratio $\eta=1$.}
\label{fig:polmol}
\end{figure}

\emph{(2) Polaron-molecule transition}.--
The stability of molecular state is challenged by the polaron state, which is characterized by particle-hole excitations above the Fermi-sea. Up to the lowest order excitation, the polaron wave function is written as:
{

\begin{align}
\left|P\right\rangle=\Big(\varphi_0 b^{\dag}_{0} +\sum_{\substack{\lambda_1\\ \lambda_2}} \sum_{\substack{\xi^a_{{\bf k},\lambda_1}>E_h \\ \xi^a_{{\bf q},\lambda_2}<E_h }} \varphi_{\cp k,\cp q}^{\lambda_1\lambda_2}b^{\dag}_{\cp q-\cp k}a^{\dag}_{\cp k,\lambda_1}a_{\cp q,\lambda_2}\Big) \left|FS\right\rangle_N,
\end{align}}
where $\lambda_i=\pm$ ($i=1,2$) denotes the two helicity branches, $\varphi_{\cp k,\cp q}^{\lambda_1\lambda_2}$ and $\varphi_0$ are the variational coefficients. Minimizing the energy functional, we have the equation for the polaron energy $E_p$:
\begin{equation}
E_p=\sum_{\substack{\lambda_2\\ \xi^a_{{\bf q},\lambda_2}<E_h}} \frac{(\beta^{\lambda_2}_{\cp q})^2}{U^{-1}-\displaystyle\sum_{\substack{\lambda_1\\ \xi^a_{{\bf k},\lambda_1}>E_h}}\frac{(\beta^{\lambda_1}_{\cp k})^2}{E_p-\epsilon^b_{\cp q -\cp k}-\xi^a_{\cp k,\lambda_1}+\xi^a_{\cp q,\lambda_2}}} .\label{eqn:polaE}
\end{equation}
Here we have considered a polaron state with zero CoM momentum~\cite{footnote}.

{In Fig.~\ref
{fig:polmol}(a), we compare the polaron energy $E_p$ with $E_M-E_h$, the molecular energy relative to a spin-orbit coupled Fermi sea of $N$ atoms, for various interaction strengths as $h$ varies.} Interestingly, the trajectories of these two energies behave differently when the number of Fermi surfaces changes from two to one as $h$ increases. Consequently, for appropriate parameters, two separate transition points can be crossed by tuning $h$ (see dash-dotted curves in Fig.~\ref{fig:polmol}(a)). In Fig.~\ref{fig:polmol}(b), we plot the polaron-molecule transition boundaries as functions of $h$. One can see clearly the non-monotonic behavior of the critical $1/a_s$, with a minimum kink appearing when the Fermi surface just touches the upper helicity branch. Thus for certain fixed interaction strengths, the system can go through two transition boundaries by changing the SOC parameter $h$, leading to an exotic reentrance of the polaron state. This reentrance phenomenon has never been found in previous studies of polaron-molecule transition in Fermi gases.

\emph{Many-body phase diagram}.--
The rich physics discussed above should also be manifest in the pairing phases and phase transitions on the many-body level. To demonstrate this, we study the pairing superfluidity of the system using the standard BCS-type mean-field theory, where the thermodynamic potential at zero temperature $\Omega=\left\langle H-\mu_a N_a-\mu_b N_b\right\rangle$ becomes~\cite{supp}:
\begin{equation}
\Omega=\sum_{{\bf k}\gamma}\theta(-E_{\cp k \gamma})E_{\cp k \gamma}
+\sum_{\bf k} (\epsilon^a_{\frac{\cp Q}{2}+\cp k,+}+\epsilon^a_{\frac{\cp Q}{2}+\cp k,-}) -V\frac{\Delta_{\cp Q}^2}{U}.
\end{equation}
Here, $\epsilon^a_{\cp k,\pm}=\epsilon^a_{\cp k}\pm \alpha k_x-\mu_a$, $E_{\cp k \gamma}(\gamma=1,2,3)$ is the quasi-particle energies, the order parameter $\Delta_{\bf Q}=U/V \sum_{\bf k} \langle b_{\frac{\cp Q}{2}-\cp k}a_{\frac{\cp Q}{2}+\cp k,\uparrow} \rangle$, $\mu_a$ ($\mu_b$) is the chemical potential of the corresponding atoms, $N_a$ ($N_b$) is the corresponding particle number, and $\theta(x)$ is the Heaviside step function. Given the chemical potentials, the ground state can be obtained by minimizing $\Omega$ with respect to ${\bf Q}$ and $\Delta_{\bf Q}$. Numerically, we find that the ground state ${\bf Q}$ for the FF pairing always lies along the $x$-direction.

\begin{figure}[tbp]
\includegraphics[width=9cm]{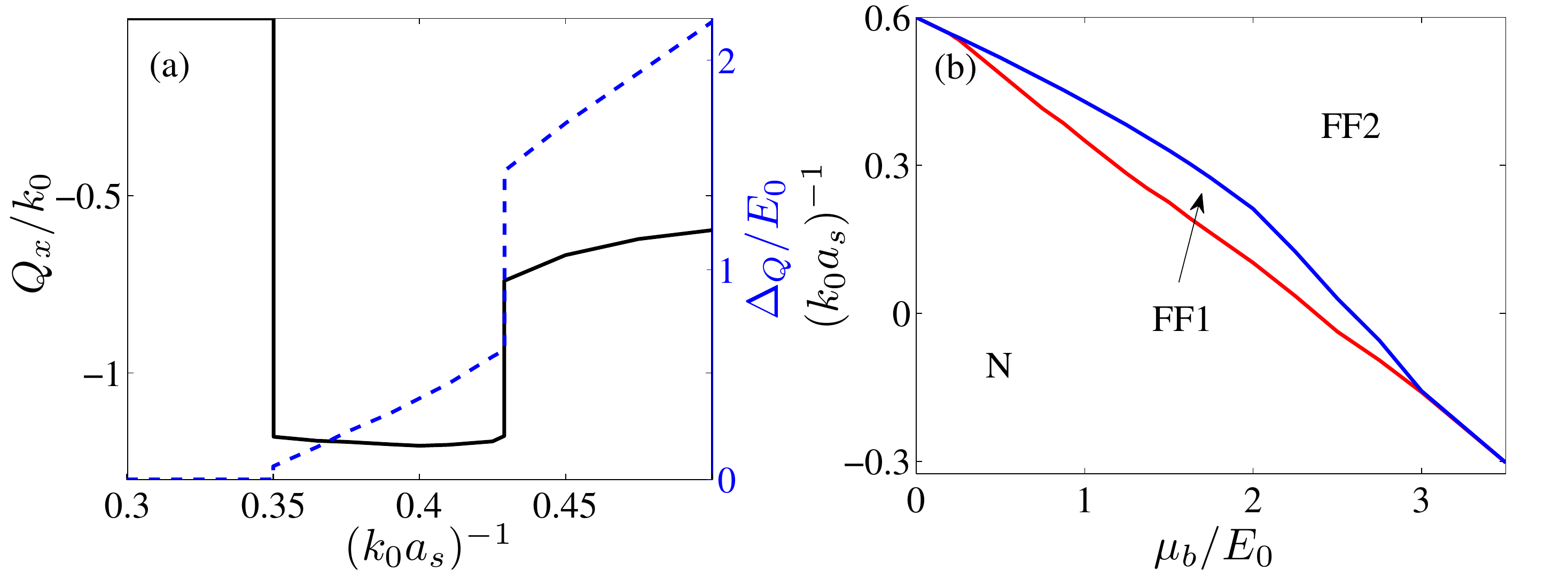}
\caption{(Color online) (a) Ground state CoM momentum $Q_x$ (solid, left $y$-axis) and pairing gap $\Delta_{\cp Q}$ (dashed,right $y$-axis) as functions of $1/(k_0a_s)$. Here we choose $(h,\mu_a,\mu_b)=(0.5, 0.75, 1)E_0$. (b) Phase diagram in the ($\mu_b/E_0, (k_0a_s)^{-1}$) plane. The red (blue) curve shows the N-FF1 (FF1-FF2) transition boundary, where both boundaries are of first order. $h$ and $\mu_a$ are the same as in (a)(b). In all cases, the mass ratio $\eta=40/6$.}
\label{fig:phase}
\end{figure}

In Fig.~\ref{fig:phase}(a), we show how the ground state $Q_x$ and $\Delta_{\cp Q}$ evolve with the interaction parameter $1/(k_0a_s)$, with other parameters ($h,\mu_{a,b}$) fixed. One sees clearly that by increasing $1/(k_0a_s)$, the system first goes through a first-order phase transition from the normal phase (N) to an FF pairing phase (FF1), and then another first-order transition to a new FF phase (FF2) with different $Q$ and $\Delta_{\cp Q}$. {
The emergence of multiple FF phases and the first-order boundaries between them are qualitatively consistent with our previous analysis on molecular states, though the actual values of $Q_x$ and $a^{-1}$ at the phase boundary are now affected by the existence of the impurity Fermi sea.} In Fig.~\ref{fig:phase}(b), we show the phase diagram when the chemical potential $\mu_b$ is further tuned. Again, over a wide region of $\mu_b/E_0\in(0.2, 3)$, the system can be tuned across all three phases (N-FF1-FF2) by adjusting the interaction strength. Hence, these exotic phases and phase transitions can be probed experimentally by sweeping the magnetic field across a Feshbach resonance of the spin-up and the impurity atoms.

\emph{Summary}.--
In summary, we have systematically studied the pairing physics in a three-component Fermi-Fermi mixture, where two constituent fermions are subject to synthetic spin-orbit coupling. We show that the ground state pairing of this system intrinsically features a finite CoM momentum, which originates from a distinctively new FF pairing mechanism in spin-orbit coupled Fermi systems.
The intriguing properties of the two-body bound state, the polaron-molecule transition and the FF paring superfluidity in this system can be explored in current cold atom experiments on $^{40}$K-$^{40}$K-$^{40}$K or $^{6}$Li-$^{40}$K-$^{40}$K mixtures. {
For example, in a $^{40}$K-$^{40}$K-$^{40}$K mixture, a Raman process with a coupling strength of $h\sim2E_r$ and a Fermi energy of $E_F\sim 4 E_r$ should lead to $\alpha k_F/E_F \sim1$ and $h/E_F\sim0.5$. Here the recoil energy $E_r$ ($E_r=E_0/4$) is typically on the order of $8.3$kHz for $^{40}$K~\cite{fermisocexp1}. Under these parameters, the polaron-molecule transition occurs at $(k_Fa_s)^{-1}\sim 1.2$. By further varying the parameters $(h, \  a_s)$, or by choosing different atomic species and adjusting their numbers, many of the interesting phenomena discussed in this work should emerge given low enough temperatures. }

\emph{Acknowledgments}.-- We thank Jing Zhou for her early contribution to this project.
This work is supported by NFRP (2011CB921200, 2011CBA00200), NNSF (60921091), NSFC (11104158,11374177,11105134,11374283), the Fundamental Research Funds for the Central Universities (WK2470000006), and the programs of Chinese Academy of Sciences.

\clearpage
\begin{widetext}
\appendix
\section{Supplementary material}

\subsection{Model Hamiltonian in the helicity basis}

Under SOC, it is more transparent to re-write the Hamiltonian in the helicity basis:
\begin{align}
H=&\sum_{\cp k,\lambda=\pm}\xi^a_{\cp k,\pm}a^{\dag}_{\cp k,\lambda}a_{\cp k,\lambda}+\sum_{\cp k}\epsilon^b_{\cp k} b^{\dag}_{\cp k}b_{\cp k}\nonumber\\
&+U\sum_{\cp k,\cp k',\cp q} \left(\beta^+_{\frac{\cp q}{2}+\cp k}a^{\dag}_{\frac{\cp q}{2}+\cp k,+}-\beta^-_{\frac{\cp q}{2}+\cp k}a^{\dag}_{\frac{\cp q}{2}+\cp k,-}\right)\nonumber\\
&\times b^{\dag}_{\frac{\cp q}{2}-\cp k}b_{\frac{\cp q}{2}-\cp k'}\left(\beta^+_{\frac{\cp q}{2}+\cp k'}a_{\frac{\cp q}{2}+\cp k',+}-\beta^-_{\frac{\cp q}{2}+\cp k'}a_{\frac{\cp q}{2}+\cp k',-}\right),\label{eqn:Hfin}
\end{align}
where the annihilation operators of the helicity branches $a^{\dag}_{\cp k,\pm}=\pm\beta_{\cp k}^{\pm}a_{\cp k,\uparrow}+\beta_{\cp k}^{\mp}a_{\cp k,\downarrow}$, with corresponding  dispersion $\xi^a_{\cp k,\pm}$. The interaction between the impurity atom and atoms in the helicity branches is thus momentum dependent.

\subsection{Tuning the actual Fermi energy}

We consider a system with a fixed total particle density, as is the case in experiments. When the total particle density is fixed, changes in the SOC parameters $h$ or $\alpha$ would affect the single particle dispersion, which would change the Fermi energy $E_h$ as well. Since it is more convenient to vary the intensity of the Raman fields experimentally, we focus on the cases where $\alpha$ is fixed, while $h$ varies. For simplicity, we only consider cases with $(\alpha k_F/E_F)^2\leq 2h/E_F$, for which there is no single-particle ground state degeneracy in the lower helicity branch. There are then two Fermi surfaces, one in each helicity branch, for $E_0>h$; and only one Fermi surface in the lower branch for $-h<E_h<h$. For given parameters, the Fermi energy $E_h$ can be determined from the total number density $n$ as $n=(1/V)\sum_{\lambda,\xi^a_{{\bf k},\lambda}<E_h}$. The contour of the Fermi surface in momentum space can be calculated from the expression $E_h=\xi_{\cp k, \pm}$, which in turn determines the range of integrations in the closed equations in the main text. A typical evolution of the Fermi surface with varying parameters is given in  Fig.~\ref{fig:E0alp05}.

\begin{figure}[h]
\includegraphics[width=8cm]{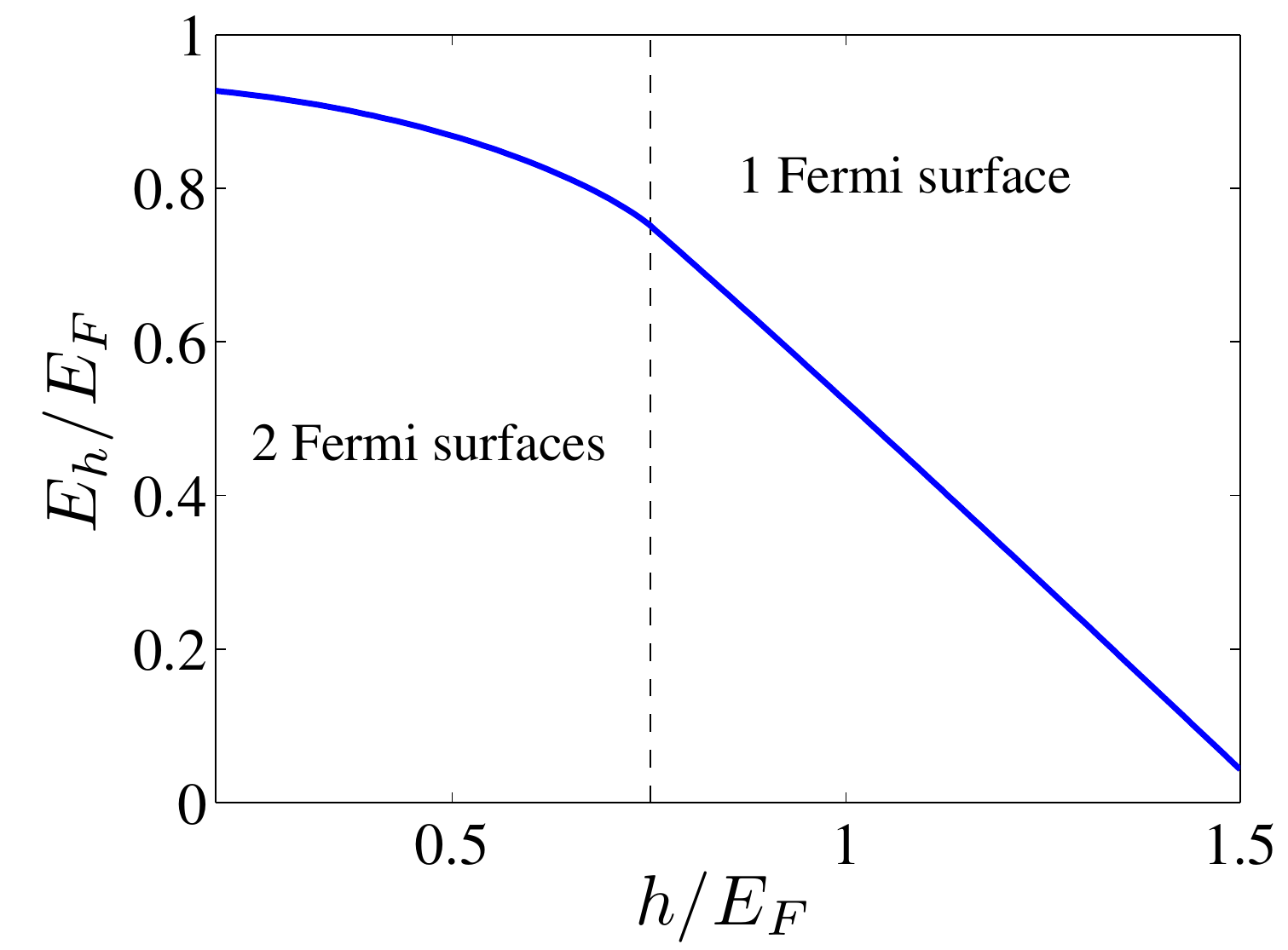}
\caption{Typical evolution of the Fermi surface with varying $h$ and a fixed $\alpha k_F/E_F=0.5$. The dashed line marks the critical Zeeman field $h/E_F\sim 0.75$, at which fermions start to populate the upper helicity branch.}
\label{fig:E0alp05}
\end{figure}

\subsection{Mean-field BCS theory}

Under the BCS mean-field treatment in the text, the effective Hamiltonian can be written in a quadratic form:
\begin{equation}
H-\mu_aN_a-\mu_bN_b=\sum_{\cp k}
\psi_{\cp k}^{\dagger}H_{\cp k}\psi_{\cp k}
+\sum_{\cp k}(\epsilon_{\frac{\cp Q}{2}+\cp k,+}^a+\epsilon_{\frac{\cp Q}{2}+\cp k,-}^a)-V\frac{\Delta_{\cp Q}^2}{U},
\end{equation}
where $H_{\bm{k}}$ is a 3$\times$3 matrix:
\begin{equation}
H_{\cp k}=\left(
  \begin{array}{ccc}
    \epsilon_{\frac{\cp Q}{2}-\cp k}^b-\mu_b & \bigtriangleup_{\cp Q}&0\\
    \bigtriangleup_{\cp Q}&-\epsilon_{\frac{\cp Q}{2}+\cp k,+}^a&-h\\
    0&-h&-\epsilon_{\frac{\cp Q}{2}+\cp k,-}^a\\
  \end{array}
\right),
\end{equation}
where we have assumed the order parameter $\Delta_{\cp Q}$ to be real.

Applying the unitary transfer on the spin basis:
\begin{equation}
\left(
  \begin{array}{ccc}
    b_{\frac{\cp Q}{2}-\cp k}^\dagger,& a_{\frac{\cp Q}{2}+\cp k,\uparrow},&a_{\frac{\cp Q}{2}+\cp k,\downarrow}\\
  \end{array}
\right)
=\left(
  \begin{array}{ccc}
    \alpha_{\cp k}^\dagger & \beta_{\cp k}&\gamma_{\cp k}\\
  \end{array}
\right)S^{\dag},
\end{equation}
$\Omega=\left\langle H-\mu_aN_a-\mu_bN_b\right\rangle$ can be diagonalized:
\begin{equation}
\Omega=\sum_{\cp k}\left(
  \begin{array}{ccc}
    \alpha_{\cp k}^\dagger & \beta_{\cp k}&\gamma_{\cp k}\\
  \end{array}
\right)
S^{\dag}H_{\cp k}S
\left(
  \begin{array}{ccc}
    \alpha_{\cp k}\\\beta_{\cp k}^\dagger\\ \gamma_{\cp k}^\dagger\\
  \end{array}
\right)
+\sum_{\cp k}(\epsilon_{\frac{\cp Q}{2}+\cp k;+}^a+\epsilon_{\frac{\cp Q}{2}+\cp k;-}^a)-V\frac{\Delta_{\cp Q}^2}{U},
\end{equation}
where $S$ is the transformation matrix containing three eigen-vectors of $H_{\bf k}$, and $\alpha_{\bf k}, \beta_{\bf k}, \gamma_{\bf k}$ are the annihilation operators for quasi-particles with energies $E_{{\bf k}\gamma}\ (\gamma=1,2,3)$. Then for the zero-temperature ground state we can get the expression of the thermodynamic potential $\Omega=\left\langle H-\mu_aN_a-\mu_bN_b\right\rangle$ as Eq.(7) in the text, where the expectation value $\left\langle\cdot\right\rangle$ is taken with respect to the vacuum state of the quasi-particles.
\end{widetext}


\begin{thebibliography}{99}
\bibitem{footnote_threshold} {$E_{th}$ is defined as the lowest scattering energy of a free impurity atom and a free atom under SOC. Correspondingly, $E_{th}=-h$ for $h/E_0>1/2$, and $E_{th}=-E_0/4-h^2/E_0$ otherwise.}
\bibitem{footnote} It is straightforward to show numerically that the ground state in the polaron sector features a very small nonzero center-of-mass momentum, and a very small decrease in energy.
\end{thebibliography}
\end{document}